# Automated, Targeted Testing of Property-Based Testing Predicates

Tim Nelson[a], Elijah Rivera[a,b], Sam Soucie[a,c], Thomas Del Vecchio[a], John Wrenn[a], and Shriram Krishnamurthi[a]

a Brown University, USA

b Massachusetts Institute of Technology, USA

c Indiana University, USA

**Abstract**

**Context** This work is based on property-based testing (PBT). PBT is an increasingly important form of software testing. Furthermore, it serves as a concrete gateway into the abstract area of formal methods. Specifically, we focus on students learning PBT methods.

**Inquiry** How well do students do at PBT? Our goal is to assess the quality of the predicates they write as part of PBT. Prior work introduced the idea of decomposing the predicate's property into a conjunction of independent subproperties. Testing the predicate against each subproperty gives a "semantic" understanding of their performance.

**Approach** The notion of independence of subproperties both seems intuitive and was an important condition in prior work. First, we show that this condition is overly restrictive and might hide valuable information: it both undercounts errors and makes it hard to capture misconceptions. Second, we introduce two forms of automation, one based on PBT tools and the other on SAT-solving, to enable testing of student predicates. Third, we compare the output of these automated tools against manually-constructed tests. Fourth, we also measure the performance of those tools. Finally, we re-assess student performance reported in prior work.

**Knowledge** We show the difficulty caused by the independent subproperty requirement. We provide insight into how to use automation effectively to assess PBT predicates. In particular, we discuss the steps we had to take to beat human performance. We also provide insight into how to make the automation work efficiently. Finally, we present a much richer account than prior work of how students did.

**Grounding** Our methods are grounded in mathematical logic. We also make use of well-understood principles of test generation from more formal specifications. This combination ensures the soundness of our work. We use standard methods to measure performance.

**Importance** As both educators and programmers, we believe PBT is a valuable tool for students to learn, and its importance will only grow as more developers appreciate its value. Effective teaching requires a clear understanding of student knowledge and progress. Our methods enable a rich and automated analysis of student performance on PBT that yields insight into their understanding and can capture misconceptions. We therefore expect these results to be valuable to educators.

## 1 Introduction

*Property-based testing* (PBT) techniques have gained mainstream attention with new libraries being published for many languages, with discussions appearing on developer blogs, social media, etc. PBT is an especially powerful technique in real-world software systems where (e.g.) the specification is still in flux, behavior is non-deterministic, or developers simply wish to continuously "mine" for bugs in the cloud.

Unfortunately, at time of writing, PBT is not commonly covered in undergraduate computer-science education. This lamentable neglect was the topic of prior work [40] in this journal (by a subset of the present authors)—henceforth PJ21—which argued that PBT could productively be taught to undergraduates in multiple settings. Teaching *effectively* requires giving accurate and focused feedback (and if that can be automated, so much the better). This raises an important question, especially in the educational setting: *how should we assess and critique the quality of property-based testing*?

**Breaking down PBT** Broadly, PBT comprises two halves: a *generator* that produces inputs to the implementation under test, and a *predicate* that determines whether the implementation's output is valid for a given input. While both pieces are vital, we take the view that in education the *predicates* are of chief importance, in part owing (as PJ21 notes) to their utility as a gateway to specification and formal methods. Also, suitable generators are often built into professional PBT frameworks; the predicates are therefore the one piece that a student is guaranteed to need to write themselves. We thus restrict our attention to the predicates students write on PBT assignments.

Since every predicate is a binary classifier on input-output pairs, a first cut at evaluating their quality might involve generating a set of pairs and measuring how accurately the predicate classifies them. But PJ21 notes that, while this could tell us how incorrect the predicates are, it would fail to give any semantic insight into the *kinds of bugs* present—a vital precondition for educational feedback.

Instead, our previous work took the view that PBT predicates are frequently (whether implicitly or explicitly) a conjunction of smaller predicates that check different fine-grained aspects of a problem. E.g., a predicate for checking sort implementations needs to confirm that the output is sorted, but also that the output is a permutation of the original input (which itself involves, at minimum, separate containment checks in each direction). If stability is specified, that must also be checked, and so on.

A formal specification of correctness can thus be usefully *decomposed* into a set of narrow subproperties that, together, are equivalent to the original. In this way, a PBT predicate can be evaluated not as a monolith but as a conjunction of fine-grained classifiers, only some of which may be erroneous. Exercising predicates separately for each subproperty yields focused insight into what went wrong, enabling partial credit and, since submissions can be clustered, easing the task of giving actionable, individualized feedback.

**Generating Useful Tests** To execute this idea, PJ21 produced a test suite for each subproperty. Each suite comprised input-output pairs that violated the corresponding subproperty *and satisfied all others* (so that a failure would unambiguously correspond to a single subproperty). Achieving this perfectly required a lack of logical implications between subproperties. PJ21 called this criterion *independence*, and relied upon it to evaluate student work. Since the goal was to understand the kinds of errors students made while learning PBT, and to assist in giving feedback, the sharp focus lent by independence made it appealing. Unfortunately, two new problems arose.

1. Independence exacerbated the incompleteness of testing: if a predicate failed only on input-output pairs wherein *multiple* subproperties failed, the error would never be discovered by a method using only single-failure pairs.
2. Independence limited use: the more subproperties we added, the more difficult it was to maintain independence. This was especially true when new subproperties corresponded to specific, pedagogically important errors (section 3).

This work corrects both concerns. Moreover, while subproperties must still be decomposed manually, this work automates the previously onerous and error-prone manual production of narrowly-focused test suites. In short, our contributions are:

- demonstrating why the independent-subproperties criterion was flawed (section 3);
- correcting these flaws while retaining the essence of the original (section 4);
- exploring multiple automated methods of test-suite generation (sections 5–7); and
- evaluating how these methods compare in error-catching power and scalability, relating these automated approaches to a human best effort, and re-assessing the results from PJ21 (section 9).

## 2 Background and Terminology

Before beginning, we quickly summarize the PBT assignments for which our students wrote predicates (because the specifics will matter later). Although PJ21 describes these in great detail, for completeness we will also summarize them here. We also need to define some terms to avoid confusion, because many are used inconsistently across contexts. Readers eager to cut to the chase can skip the latter on first read.

### 2.1 Pedagogic Context and Assignments under Test

This work is set in the context of, and evaluated on, a trio of PBT assignments from two different classes at Brown University. In all three problems, students write a full PBT solution: both a *generator* and a *predicate* (denoted, respectively, `generate-input` and `is-valid` by PJ21). Here, as in PJ21, we focus solely on the predicates that students wrote, since we are interested in how well they expressed correctness properties.

In each course, submitted predicates are graded using suites of input-output pairs, which effectively exercise the notion of predicates as classifiers of good and bad behavior. Thus, one facet of our work here lies in augmenting these preexisting suites in a directed, disciplined way; we evaluate the effectiveness of this in section 9. All courses, assignments, and corresponding datasets used here are the same as those from PJ21 [40, section 3].

**Courses** *Accelerated Intro:* The first two problems (Sortacle and Matcher) appear in an advanced introductory Computer Science course covering functional programming, basic algorithms, and asymptotic analysis. Programming assignments used Pyret, a functional language that shares many features with ML. More than 90 percent of the students are in the first semester at Brown. To gain entry to the course, students must pass a series of summertime assessments; most students thus have some prior experience with programming, albeit usually not in a functional style. Depending on year, the course had from **64** to **88** students. PBT was largely covered in the text of the assignments, along with a brief in-class discussion, and is reinforced by the class's emphasis on (and grading of) traditional unit testing.

*Applied Logic:* The third problem (Toposortacle) appears at the beginning of a tools-focused introductory formal-methods course. The course's primary focus is on modeling and reasoning about systems using tools like Alloy [11], Spin [10], and Dafny [15]. PBT was introduced early, via an in-class lecture, as a bridge from traditional testing (which all students have done) to formal specification (which few have seen). Most students are second and third year undergraduates. Depending on year, the course hosted between **92** and **147** students.

**Assignment: Sortacle** Students write a predicate for testing purported implementations of a sorting algorithm. Both inputs and outputs correspond to lists of PERSON records, where a PERSON comprises an age (the key to be sorted on) and a name (additional information). The specification does *not* stipulate that the sort is stable. For Sortacle, students' predicates had the type signature: `List<Person>, List<Person> -> Bool`. Over three years, we collected 205 student predicates for Sortacle.

Write-ups: https://cs19.cs.brown.edu/{2017, 2018, 2019}/sortaclesortacle.html

**Assignment: Matcher** Students write a predicate for testing implementations of the stable-marriage problem, cast as matching prospective employees with companies. Inputs are pairs of lists of lists of numbers denoting, for both people and companies, their preferences (indexes serve as identities for both). For Matcher, students' predicates had the type signature: `List<List<Number>>, List<List<Number>>, Set<(Number,Number)> -> Bool` Over three years, we collected 200 student predicates for Matcher.

Write-ups: https://cs19.cs.brown.edu/{2017, 2018, 2019}/oracleoracle.html

**Assignment: Toposortacle** Students write a predicate for testing an implementation of topological sort. Inputs comprise lists of pairs of vertices (the partial order to be obeyed). The input covers all vertices—i.e., the domain is fully given and no additional vertices need to be considered. Outputs consist of a list of vertices. For Toposortacle, students' predicates had the type signature: `List<(Number, Number)>, List<Number> -> Bool`. Over two years, we collected 237 student predicates for Toposortacle.

Write-ups: https://cs.brown.edu/courses/cs195y/{2018, 2019}/historical/oracle.pdf

## 2.2 Terminology

A *problem* $\mathscr{P}$ is a mathematical object defining a relation between elements of an *input* type and elements of an *output* type. Often, problems are expressed informally but, if well-specified, can be understood precisely. We might informally write "take the square root of the input" to mean the problem relation $\{(i,o) \in \mathbb{R} \times \mathbb{R} : o^2 = i\}$. If there are multiple valid outputs for a given input (as in this example: both $1^2 = 1$ and $(-1)^2 = 1$), we call the problem *relational*. Relational problems are legion in computer science: sorting, change-making, nearly every optimization problem or graph problem, etc. all have inputs with multiple correct answers.

An *implementation* for a problem is a program that maps elements $i$ of the problem's input type to elements $o$ of its output type. Crucially, an implementation may not be correct; it is *correct* exactly when for every $i$, $(i,o)$ is in the problem's relation. Because of this uncertainty, *testing* implementations is a vital and pervasive challenge in programming. We have argued [40] in the past that relational problems pose a special challenge in testing and pose an obvious application area for PBT.

Students write PBT *predicates* that accept an *input-output pair* $(i,o)$ and return a boolean indicating whether that pair is in the problem's relation. It would be reasonable to ask, with some trepidation, whether a PBT predicate for testing implementations of $\mathscr{P}$ is itself an implementation for the meta-problem of recognizing valid input-output pairs for $\mathscr{P}$. The answer is yes, which is why this line of work exists. (With one exception in section 7, we pass over the question of testing the tester of the tester....)

Since the base problem's relation is over pairs $(i,o)$, the predicate meta-problem's relation is over pairs whose input is $(i,o)$ and the output is a boolean $b$, formally, $\{((i,o),b) : b \iff (i,o) \in \mathscr{P}\}$. When speaking of generating or evaluating a predicate on input-output pairs, we will usually have a $b$ in mind that is clear from context (i.e., whether we are producing, or running, positive or negative tests).

A *property* $P$ is a finite logical formula over $(i,o)$ pairs. We assume that every problem under consideration can be soundly captured by such a formula, and will thus sometimes refer to a problem and its exact correctness property interchangeably. Properties may be *satisfied*, or not, by $(i,o)$ pairs. We will sometimes use similar notation for the same idea, speaking of (e.g.) pairs *obeying* or *disobeying* a property.

A *subproperty* for a problem $\mathscr{P}$ is possibly an underspecification of the problem: it is implied by $\mathscr{P}$ but does not necessarily imply $\mathscr{P}$. Any property (and thus any problem) may be *decomposed* into a set of subproperties $S$ whose conjunction is logically equivalent to $\mathscr{P}$. We say that $S$ is *independent* if no proper subset of $S$ is equivalent to $S$—that is, if every subproperty in $S$ excludes at least one input-output pair that the others do not.

A *bucket* for a set of subproperties $S$ is a member of the power set of $S$. Where context is clear, we will omit naming $S$. A *test suite* is a non-empty, finite set of input-output pairs. A suite is said to *concretize* a bucket $b$ for $S$ if every subproperty in $b$ evaluates to true on (is satisfied by) every member of the suite and every subproperty in $(S \setminus b)$ evaluates to false on (is not satisfied by) every member of the suite.

**Table 1** Original subproperty decomposition for Toposortacle, in terms of an input-output pair $(\mathbf{I}, \mathbf{O})$, along with an example test disobeying each property (and only it). The *set* operator denotes the list-to-set coercion, and *len* is the length of the list. $\mathbf{I}^*$ is the transitive closure of the input, i.e., the partial order induced by the edges provided. All example test pairs witness the failure of their corresponding subproperty without violating any others; this is possible because of independence.

Original Toposortacle Properties

| | | |
|---|---|---|
| P1 | No-New | $\forall\, n \in set(\mathbf{O}) : \exists m : (m,n) \in \mathbf{I} \vee (n,m) \in \mathbf{I}$ |
| P2 | None-Dropped | $\forall\,(m,n) \in \mathbf{I} : m \in set(\mathbf{O}) \wedge n \in set(\mathbf{O})$ |
| P3 | Uniqueness | $\forall\, 0 \le i_1 < i_2 < len(\mathbf{O}) : \mathbf{O}[i_1] \ne \mathbf{O}[i_2]$ |
| P4 | Sortedness | $\forall\, 0 \le i_1 < i_2 < len(\mathbf{O}) : (\mathbf{O}[i_2], \mathbf{O}[i_1]) \notin \mathbf{I}^*$ |

Example Input-Output Pairs

| | |
|---|---|
| No-New | $([(1,2),(2,3)],[1,2,3,4])$ |
| None-Dropped | $([(1,2),(2,3)],[1,2])$ |
| Uniqueness | $([(1,2),(2,3)],[1,2,3,3])$ |
| Sortedness | $([(1,2),(2,3)],[3,1,2])$ |

## 3 Why Not Independence?

We choose Toposortacle as a running example, although we will later evaluate with all three assignments. This is because Toposortacle is the assignment whose properties are easiest to express concisely (there are no concerns about records or duplicate entries, as in Sortacle, or preference-orderings, as in Stable Matching).

Table 1 lists an initial decomposition of correctness for Toposortacle. This decomposition is independent: for every subproperty $P$, the figure shows an input-output pair that makes $P$ false but all other listed subproperties true. It is identical to the decomposition used in PJ21 [40, section 4.3], except that it divides the "Same-Elements" subproperty into two separate statements (No-New and None-Dropped). This is consistent with the overall goal of decomposition, and does not affect independence. The third property, Uniqueness, is necessary because students were told that the input covered all possible vertices.

At a high level, this decomposition is reasonable and helpful: tests like those in table 1 can detect failures of specific subproperties. But, as briefly mentioned in section 1, there are issues lurking beneath the independent decomposition.

### 3.1 Bugs May Escape Notice

Focusing on tests that violate a single subproperty yields extremely clear feedback about what went wrong—which is why PJ21 chose independence. However, focusing on *only* those can miss bugs that only fail when multiple subproperties are violated. Indeed, we find that students actually make such errors (section 9)!

### 3.2 Independence Is Challenging to Maintain

When modeling Toposortacle in Alloy [11], we had originally defined Sortedness as: $\forall (m,n) \in I : \text{idxOf}(O,m) < \text{idxOf}(O,n)$. That is, we said that for every pair $(m,n)$ in the input, $m$ occurs before $n$ in the output. But this property relies on None-Dropped (or else there may be no "index of" an element) and Uniqueness (or else we must speak of the *first* index of an element). This pattern repeated as we worked: often a natural way to express a subproperty is dependent on other subproperties, and producing an independent variant is extra work. Even then, without a careful regime of formal checking, it is easy to make mistakes.

Indeed, we only noticed this flaw in the original property because we were able to compare corresponding generated tests against a canonical test suite for the problem that had been carefully crafted, over multiple years, by course staff.

### 3.3 Useful Decompositions Often Cannot Be Independent

There is an error students commonly make on Toposortacle: they check the equivalence of the input and output vertex-sets by comparing only their *size*, rather than their contents. From manual inspection of these flawed predicates, it appears that their authors either have not carefully thought through what they need to express, or else are prey to some deeper misconception about sets. In either case, there is an issue that needs to be addressed with immediate and actionable feedback.

To detect instances of this error, we could create a new subproperty that characterizes what these submissions get right: that *the output must contain the same number of vertices as appear in the input*. We can then search for predicates that accurately check the new property yet fail to correctly check either None-Dropped or No-New. In such cases, we know that the predicate is definitely flawed, but because it catches violations of the new subproperty, it (likely) only uses list length or some variation.

We call these new properties *error subproperties* because they are created to help identify a specific error, usually when an existing subproperty is only imperfectly checked. Unfortunately, this idea inexorably leads to non-independent subproperty sets. Consider the following (new) subproperty of Toposortacle, which states that the *number* of vertices mentioned in the input must equal the *length* of the output list:

P5 Same-Num-Vertices $|\{n : \exists m : (n,m) \in I \vee (m,n) \in I\}| = len(O)$

Same-Num-Vertices is implied (figure 1) by a conjunction of two existing subproperties from table 1: No-New and None-Dropped. This is only to be expected, since our strategy hinges on the error subproperty being "more permissive" than some others. The contrapositive of this implication says that whenever Same-Num-Vertices fails, either No-New or None-Dropped must also fail. Suddenly, we no longer have the luxury of generating test suites for buckets where only one subproperty is negated.

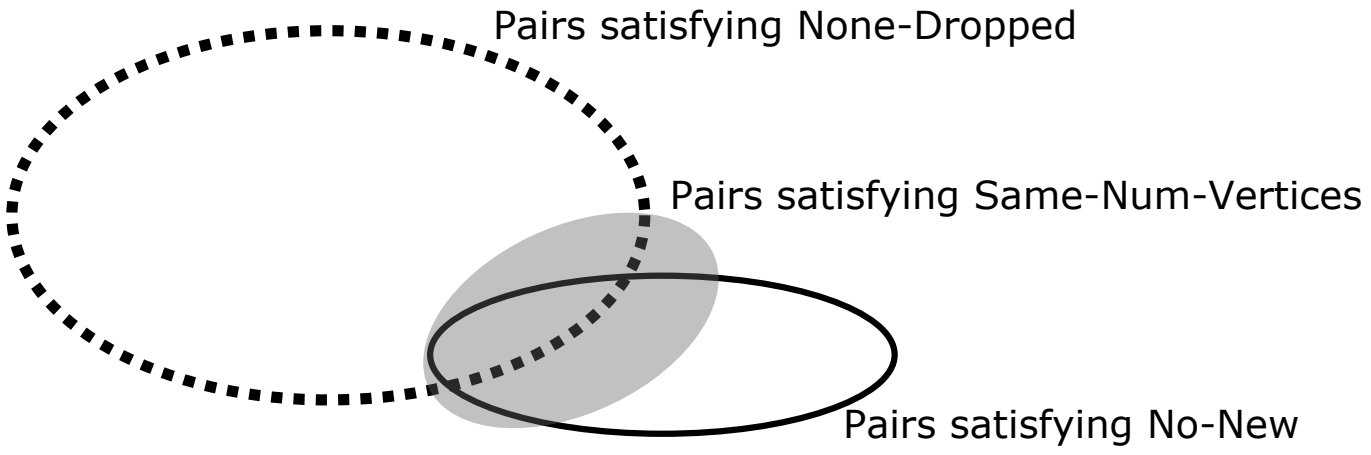

**Figure 1** Illustration of the relationship between None-Dropped, No-New, and Same-Num-Vertices.

## 4 Productively Weakening Independence

The solution we adopt is to admit any decomposition $S$ of the problem, and consider *in principle* all buckets $b$ in the power set of $S$. Since every subproperty is a formula over input-output pairs, each $b$ induces a formula $(\bigwedge_{p\in b} p) \wedge (\bigwedge_{p\in(S\setminus b)} \neg p)$, that is satisfied by exactly those pairs witnessing the subproperty failures and successes dictated by $b$.

To generate these input-output pairs, we use two distinct methods: one built atop a well-used PBT library, and another that leverages a SAT-solver. But since any such generator is always run with respect to buckets, and seeks tests satisfying that bucket's criteria, it is well worth considering which buckets to generate tests for: some may be of special interest, and others will be contradictory and impossible to concretize. We illustrate this with a continuation of the error-property example of section 3.

### 4.1 A Worked Example

Consider two subproperties for Toposortacle: the error property Same-Num-Vertices from section 3, and the conjunction of No-New and None-Dropped, which we denote as Same-Elements. (We omit other subproperties and combine these two for simplicity in presentation.) Together, these two subproperties lead to $2^2 = 4$ buckets:

(B1) {Same-Elements = False, Same-Num-Vertices = False};

(B2) {Same-Elements = True, Same-Num-Vertices = False};

(B3) {Same-Elements = False, Same-Num-Vertices = True}; and

(B4) {Same-Elements = True, Same-Num-Vertices = True}.

The fourth bucket, B4, would yield only positive tests in this example, so we disregard it here. Because Same-Elements implies Same-Num-Vertices, bucket B2 cannot be populated. The remaining two buckets can be concretized into test suites. Some example input-output pairs might be:

For B1: ([(1,2), (2,3)], [1, 2]);

For B3: ([(1,2), (2,3)], [1,2,2]).

Now if a student's predicate:

- catches both test suites, no mistakes are found;
- catches neither test suite, they may have neglected Same-Elements entirely;

- catches B1 but not B3, it is highly probable that they have used list-length to check the property; and if it
- catches B3 but not B1, an error has been identified of some unknown new shape — we are prompted to examine the student's predicate more closely.

Again, despite the subproperty set not being independent—indeed *because* of it—we can use this technique to distinguish between predicates with a specific kind of semantic flaw and those with some other, arbitrary error. Without the error subproperty to act as a marker, we would only be able to tell that some portion of SAME-ELEMENTS has been neglected, and automated feedback would be far poorer for it.

As manual inspection reveals additional errors, more subproperties can be added to identify each—much like the discovery of a bug leads immediately to a regression test. And by adding new error subproperties, rather than specific input-output pairs, variants of the same broad error might be detected.

### 4.2 Bucket Enumeration: Maximizing Focus vs. Error Detection

We sketch two extremes on the spectrum of bucket-enumeration strategies, one that allows for error subproperties while retaining the flavor of independence (and potentially missing some errors), and one that concretizes all possible buckets.

**Maximizing Focus** For student feedback, the most valuable buckets may still be those that focus on a specific failing subproperty. In these, only a minimum set of subproperties evaluate to false: the property being focused on, and those needed to satisfy implications between subproperties.

But how to build such focused buckets? The key lies in taking advantage of *a priori* structure: if it is known in advance that property $B$ implies property $A$, then we know that any concretizable bucket in which $A$ is false must make $B$ false as well. In terms of our running example, any bucket that makes SAME-NUM-VERTICES false must falsify either NO-NEW or NONE-DROPPED.

A sketch of this approach proceeds as follows: given a subproperty of interest, start with the bucket where only $A$ is false (the maximally focused bucket for $A$, as in PJ21). Whenever $A$ occurs on the right-hand side of an implication $I$, build the set of buckets that negate new subproperties as required by the left-hand side of $I$. (We say "set of buckets" here to account for implications like NO-NEW $\land$ NONE-DROPPED $\implies$ SAME-NUM-VERTICES.) Recur the process on every such bucket produced, and continue to fixpoint. This final set of buckets all falsify $A$, and the set contains no buckets without a provenance constructed from known implications. This is as close to the original notion of independence as can be retained in the presence of error subproperties and other sources of implication.

**Generating the Full Power Set** Generating suites for *all buckets*, that is, for the full power set of subproperties, may still have utility. Only allowing a minimal number of subproperty failures per suite may prevent detection of some bad predicates: e.g., a predicate could be crafted to be incorrect only on input-output pairs that disobey many different subproperties.

In principle, all errors could be detected by generating enough tests that violate the original, non-decomposed correctness property. However, without some measure in place to force these tests to cover each combination of subproperties, test generation could still neglect important buckets. The most costly but also most exacting approach is to concretize all possible buckets in the power set of subproperties. As we report in section 9, this strategy did find some errors that a maximally focused strategy missed.

### 4.3 Taking Stock: Now What?

At this point, it may appear that the interesting work is over: there are dozens of mature PBT libraries, and even the most intricate lattice of subproperties could be encoded in any of these, with or without independence. However, there is a sort of type error impeding the obvious path. PBT generators produce values, and are quite good at providing high-quality random distributions and input-space coverage. These values are then *passed* to properties and implementations under test. The crucial difference is that we are interested in generating only input-output pairs that *satisfy* bucket formulas (which are conjunctions of subproperties or their negations). Consequently, to productively use PBT generators, we should expect to filter each value produced and discard what fails to satisfy the required bucket(s).

Moreover, PBT libraries tend to focus on (and do quite well at) finding a single counterexample that witnesses property failure. We would prefer to obtain *many* such examples at once; a larger number of tests per bucket increases the likelihood of finding related errors—we detail just such a situation in section 8. There are also important meta-questions that PBT libraries are less suited to answering than other tools, such as checking our conjectures about when one subproperty implies another.

In addition, we have one attribute that most applications of PBT do not: staff-crafted test suites for each assignment. These provide a target for automated approaches to beat, and a progress bar until we get there—something most PBT approaches lack. This in turn creates an opportunity for evaluation (section 9).

The rest of this paper therefore describes the challenges we faced using PBT to generate tests for PBT, the engineering effort involved, the introduction of a supporting approach based on SAT-solvers, and the lessons learned in the process.

## 5 PBT-Library-Based Test Generation

Our first approach to test generation uses Hypothesis [17], a popular PBT library for Python. Figure 2 sketches the architecture: a *bucket enumerator* produces a stream of buckets under consideration (often an enumeration of the power set of subproperties, minus those we know to be contradictory). For every such bucket, a set of *subproperty functions* for the problem are combined into the corresponding conjunction (the *checker*). A problem-specific *input-output pair generator* is then repeatedly invoked and the returned pair is checked; if it satisfies the bucket, it is added to the bucket's test suite. The generator is invoked until the desired test suite size is reached. We briefly detail the ideas behind each, and then enumerate some challenges we encountered.

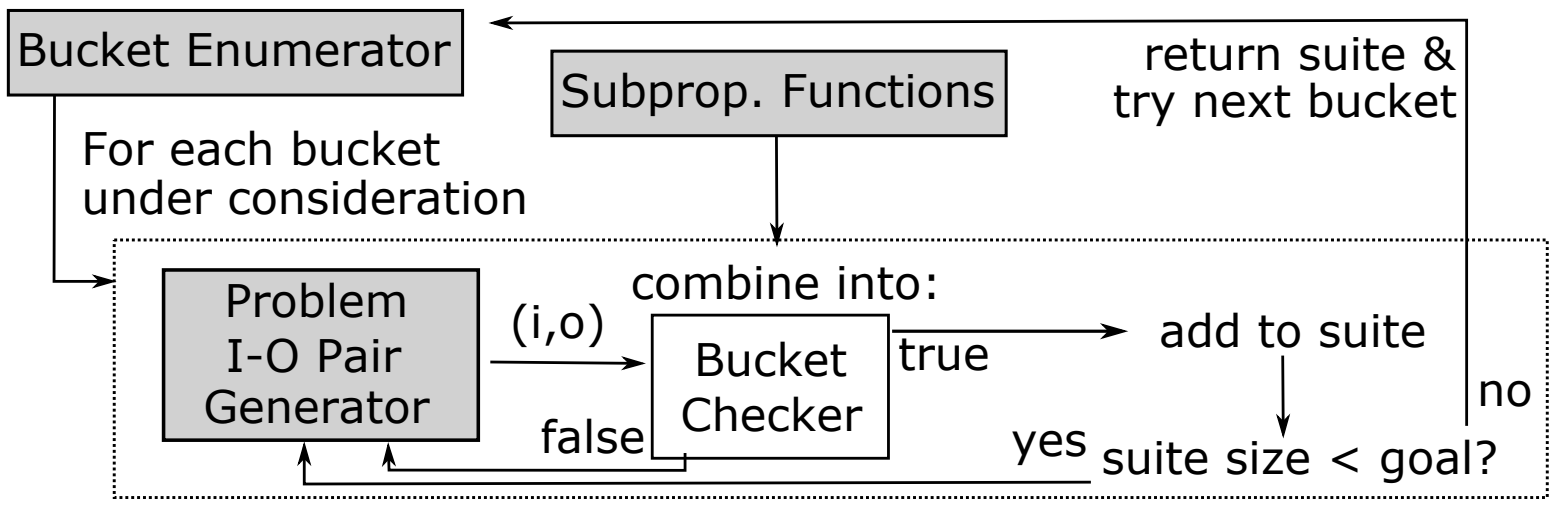

**Figure 2** Hypothesis-based test generator layout. Shaded boxes indicate components authored per-problem. The region within the dotted box is executed per-bucket, as dictated by the enumerator. Subproperties remain constant, but the bucket checker negates the appropriate combination of subproperties per bucket.

**Subproperty Functions** For each subproperty in the problem's decomposition, we write a Python function that checks this subproperty on a given input-output pair. Notice that this function is an expert-created rewrite for the subproperty in question. Since it is a Python function, not a logical specification, it must describe how to check the subproperty operationally. This is often operationally straightforward, but can involve implementing helper functions, e.g., the `transitive_closure` function so vital in checking the ordering subproperty for Toposortacle:

```
def p4(dag: DAG, output: Output) -> bool:
    partial_order = transitive_closure(dag.original_tuples)
    for i, ele1 in enumerate(output):
        for j, ele2 in enumerate(output[i+1:], start=i+1):
            if (ele2, ele1) in partial_order:
                return False
    return True
```

**I-O Pair Generator** Hypothesis has built-in generators (called "strategies") for all of the core types in Python, with much customization possible. While these sufficed in most cases, we did find it necessary to write our own strategy (listing 1) for Toposortacle, since its inputs must be directed acyclic graphs. To do this, the generator obtains a random ordering on vertices and then samples from edges that respect this ordering.

Hypothesis generators tend toward smaller values by default. This applies to not only numerics, but also the length of lists. Our first-cut implementation was therefore biased toward *trivial* tests—e.g., tests using an empty input or output list. While these tests can be useful, a suite built entirely of such tests would not be robust. Consequently, we built two variations of each generator: one that allowed trivial examples (which we invoked only once per bucket) and another that enforced a minimum list size of 3. The trivial and non-trivial generators differ only in the `min_value` parameter (listing 1).

**Bucket Enumerator** Since each bucket corresponds to a separate test suite, we must re-run generation for all buckets under consideration. The set of buckets to consider may be usefully restricted by using known implications between subproperties, or the user may opt to generate the full power set of buckets.

**Listing 1** Hypothesis generator for Toposortacle. Minimum list-length is adjusted via the min_size parameter to the standard lists generator. The trivial generator allowed a 0 length; we otherwise required an output-list length of at least 3 and at least 2 edges per input ordering. Our maximum list length was 10 for all generators.

```
def io_pair_strat_comp(draw):
    n = integers(min_value=MIN_N, max_value=MAX_N) # how many different nodes to draw edges from
    node_list = [str(i) for i in range(n)]

    # allow for nodes in output which were never used in the input
    output_node_list = [str(i) for i in range(n+1)] # 3 <= len(output_list) <= 10
    output_strat = lists(sampled_from(output_node_list), min_size=MIN_OUT, max_size=MAX_OUT)

    ordering = draw(permutations(node_list)) # Randomly permute our node list
    possible_edges = list(combinations(ordering, 2)) # all possible forward edges in ordering
    # Select a list of _distinct_ edges from possible_edges. (edges_strat)
    edges_strat = lists(sampled_from(possible_edges), min_size=MIN_DAG_EDGES, unique=True)
    dag_strat = builds(DAG, edge_tuples=edges_strat) # Pass these edges to our DAG constructor
    return draw(tuples(dag_strat, output_strat))
```

If unsatisfiable buckets are not ruled out in advance, generation can continue fruitlessly for multiple hours on a single bucket. We therefore limit the search to 20,000 candidates before moving on to the next bucket, which takes roughly 40 minutes on a laptop. We cache this information to enable skipping these buckets on subsequent runs.

**Qualitative Experience and Effort** Our efforts were aided by various features of Hypothesis (some undocumented). For instance, Hypothesis provides a .find() function that takes a function and an input generation strategy, and searches for an example that makes the given function produce true.

Still, we sometimes felt as if we and Hypothesis were working at cross-purposes. In addition to the default minimization behavior already mentioned, we encountered another issue. Because we seek multiple tests per bucket, we invoke .find() multiple times. But, by default, Hypothesis attempts to remember and reuse examples that it has previously produced! We therefore disallowed hypothesis from repeating solutions to achieve reasonable results.

The space of valid examples on all our problems was sparse enough that we needed to greatly increase the number of examples that .find() would generate and check before failing (from 100 to the 20,000 previously mentioned). A major concern is that Hypothesis cannot determine with perfect accuracy when a bucket is *not* concretizable; there is always the possibility that corresponding tests are merely underrepresented in the search space. This is a primary reason for our adding a second approach, based on SAT-solving, that is able to (far more quickly) make exhaustive claims.

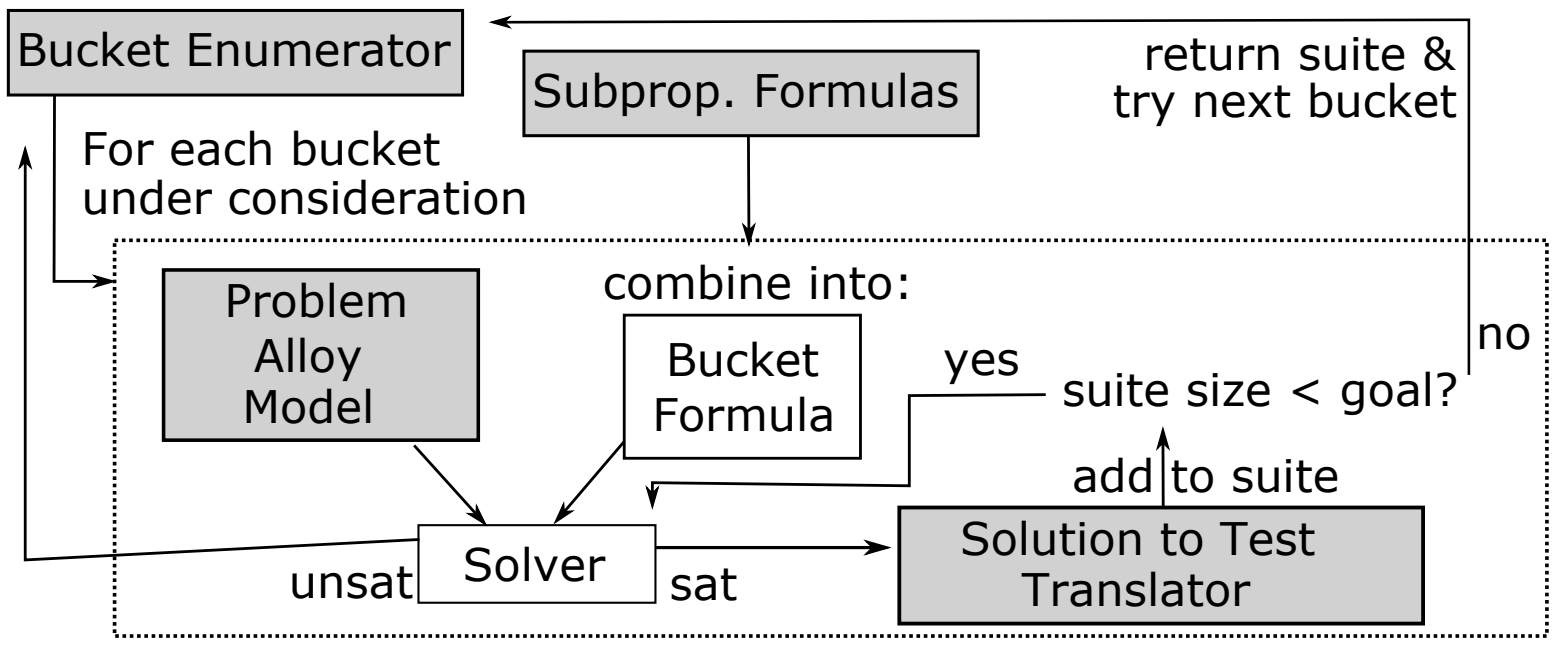

**Figure 3** SAT-based test generator layout. Shaded boxes indicate components that must be authored per-problem. The region within the dotted box is executed per-bucket, as dictated by the enumerator. Subproperty formulas remain constant, but each bucket formula negates the appropriate combination of subproperties per bucket.

## 6 SAT-Solver-Based Test Generation

Solvers have long been successfully used for test generation, even in property-focused ways [21, 22], so we decided to compare a solver-based approach to PBT. Using SAT does more than just add another test generator: it lets us rapidly explore different subproperty variants, check whether certain subproperties overlap, and much else that leverages the satisfiability perspective. Figure 3 shows the corresponding system. The overall layout is similar to our Hypothesis-backed procedure (figure 2): a Racket script imports the problem specification and the printer module, executes a search for each bucket under consideration, and outputs the results.

**Background** We use the Pardinus [4] solver, an extension of the Alloy [11] analyzer's core engine, Kodkod [38]. Alloy is a lightweight formal methods [12] tool that accepts a set of relational constraints and produces solutions to them. It has been widely used to model, verify, and refine domains like networks [18, 24, 25, 29], web security [1], UML diagrams [19, 20], source version control [28], distributed systems [41], and the core algorithms used by civil engineers in predicting storm-surge [8]. We used the Alloy toolchain not only because we were familiar users, but because the relational nature of Alloy's language eased encoding relationships in our properties. For instance, the same ordering subproperty for Toposortacle that we showed for Hypothesis is:

```
all i, j : inds[output.outR] | i < j implies { output.outR[j]->output.outR[i] !in PO[input] }
```

i.e., for all indexes $i, j$ in the output, if $i < j$ then the tuple formed by the elements at indexes $j$ and $i$ cannot be related by the input's partial order. (Alloy uses -> for cross product.) The partial-order function PO is defined as:

```
^{n2: Char, m2: Char | some pair : elems[input.inR] | pair.f = n2 and pair.s = m2}
```

where ^ is the transitive-closure operator.

**Specification and Modeling** Broadly, our tool needs only two components:

1. an Alloy specification of the problem domain and each subproperty; and
2. a Racket module that maps between the solver domain and concrete tests.

The latter component includes information such as which Alloy types correspond to test inputs and outputs, bounds (like the maximum list length), and a function that converts the solver's relational solutions into test-case strings. Usually this amounts to just a `for` loop or a functional `map`, along with some string manipulation. E.g., the solver tuple $(1,2,3)$ might become the Python list `[1,2,3]`.

Alloy specifications comprise type declarations and relational constraints. Every specification defines preconditions on the space of tests. As an example, the Toposortacle model forbids vertex-pairs with identical values:

```
sig PairChar { f: one Char, s: one Char }
all pc1, pc2: PairChar | {
  pc1 != pc2 implies {pc1.f != pc2.f or pc1.s != pc2.s} }
```

**Qualitative Experience and Effort** While Hypothesis already had a well-engineered link to Python that enabled scripting of bucket enumeration, similar scaffolding between Racket and Pardinus required additional (largely one-time) effort. Converting solutions from Pardinus was also somewhat more involved: we needed to convert them from raw relational form to Racket data structures and then to text files; output from Hypothesis was already in Python. We note here additional challenges we encountered.

*Avoiding Duplicate Tests:* In principle, solver-based approaches can easily give a stream of distinct results: given a solution, simply add a constraint that excludes it and continue the search. In practice, there are complexities. Alloy and Pardinus use a technique called Skolemization that converts existentially quantified variables into a new constant symbol to find a value for: e.g., the formula `some x: Int | x < 5` might become `c < 5`, where `c` is the fresh top-level constant. This can improve performance, but also lets every solution provide a witness for how the existential formula was satisfied. Consequently, multiple solutions might now differ only in the value of these new constants: e.g., two otherwise-identical solutions that assign 1 vs. 2 to `c`.

In our case, Skolemization at first resulted in multiple identical tests being produced for many buckets, since these fresh variables were unimportant for translating a solution into a test. This specific problem can be easily fixed by turning off Skolemization in the solver options. Other, similar issues arise from symmetries between solutions. Pardinus has built-in symmetry detection and elimination [36], but it is not perfect. This is a problem because, while actual duplicates can easily be filtered out and regenerated, isomorphic solutions are harder to detect and still lead to reduced test diversity. We found ourselves making ad-hoc changes to manually exclude solutions that produced identical tests, but were not always completely successful.

*Solver Enumerations:* Our approach relies on a boolean solver to enumerate tests. Thus, the quality of each suite it produces depends on the solver's internal search process. Different solvers will often produce differing results. Moreover, while "off-the-shelf" solvers are robust, they tend to be optimized for performance rather than diversifying outputs—the solver has no innate reason to value any particular input-output pair over another. We tried varying solvers, but found the differences (both in terms of runtime and test quality) were largely inconsequential in the aggregate. We also tried modifying the solver to follow a custom enumeration strategy (maximizing the Hamming distance from previous solutions) but likewise found the difference to be of little impact. We therefore opted to use Kodkod's default boolean solver option, SAT4J [2], leaving more intricate enumeration strategies as future work.

*Bounds:* In general, finding solutions to Alloy-style constraints is undecidable; solvers thus require users to provide a size bound on each type in the specification. For Toposortacle, this would include (e.g.) the maximum number of vertex tuples or the maximum list length. While this means the search process is incomplete, this is an issue for other test-generation techniques too. Furthermore, Jackson's [11, p. 15] *small-scope hypothesis* suggests that useful test cases are often small.

Perhaps unsurprisingly, these bounds were a major source of frustration. In principle, we needed only cap a problem-specific high-level variable (like the number of records or vertices), but bounds interact in subtle ways. An example of this phenomenon (and by no means the most complex) is the way the bound on `Int` (which sets the bitwidth of integer variables in the solver) can interact with lists. The most natural way to model finite lists in Pardinus is as a relation from integers to list elements (with some added constraints to enforce 0-indexing, contiguity, etc.). This means that even if there are (say) declared to be 5 possible list cells in a solution, lists cannot actually have that length unless the bitwidth is high enough to allow 5 distinct indexes.

While many of these issues are well-known to experienced Alloy users, they can still take time and effort to eliminate. Moreover, such issues are easy to introduce when trying to optimize performance at the Pardinus level. Fortunately, cross-validation with the Hypothesis engine (section 7) was a great help in discovering these problems, and we were often able to use the built-in core extraction features of Pardinus to help track the issue (using techniques suggested by Torlak, Chang, and Jackson [37]).

## 7 Comparison and Mutual Advantages

Both Hypothesis and the SAT-solver contributed uniquely to this work, both in terms of test quality and as a sanity check on the other's results. Neither was a clear winner; each had its own advantages. Section 9 gives a numeric report on performance and testing power, but first some qualitative comparison is called for.

The key algorithmic difference is that Hypothesis uses programmatic generators and randomness to produce tests which we then filter for suitability. In contrast, a SAT-solver's search can be more "aware" of acceptance criteria during its search. In effect, Hypothesis must operationalize what the solver is doing for free, since the generator and the property combine into one formula set. While Hypothesis does provide a "targeted example generation" feature, based on the work of Löscher and Sagonas [16], at time of writing this feature is experimental, and we leave investigating this avenue (which requires a heuristic function that approximates the quality of solutions) for future work.

Writing new properties and adopting new domains would likely be less effort in Hypothesis, since the library is able to lean on existing features of Python. But the modeling burden in Alloy is not as heavy as it may seem: one can re-use existing models (as we did) for lists, DAGs, binary trees, etc. In all, we believe adding new problems and decompositions to be within the scope of a final project in an introductory formal methods or software-engineering class.

**Cross-Validation of Results** It was easy to make mistakes in this work (as noted) and not always easy to discover them. We continuously compared the Hypothesis and Pardinus buckets to look for errors, and found several discrepancies throughout development: subtle bounds issues in our Alloy model, bad assumptions in Hypothesis, etc. We take some pleasure in noting that this cross-validation was itself related to PBT: it was a variant of model-based testing, where (e.g.) the correctness criterion for Hypothesis were given by Pardinus. This work can thus be said to employ a sort of:

- model-based testing of
  - subproperty-based test generation for evaluating
    - property-based testing predicates that
      - test programs.

## 8 Qualitative Experience: A Case Study

While performing a preliminary run of our generated tests, we encountered an interesting student solution to Toposortacle. The predicate was failing *one test* (the 9th of 10) produced by Hypothesis in *exactly one bucket*. Aside from this one flaw, the predicate passed the rest of the Hypothesis test suite, the entire Solver test suite, the hand-curated TA test suite, and the test suite from PJ21. The test was fairly complex, and was one that we had prevented Hypothesis from shrinking:

```
in = [('2', '3'), ('1', '3'), ('2', '1'), ('1', '0')]
out = ['2', '1', '3', '3', '0', '4', '0', '3', '1', '4']
```

This test fails three subproperties of Toposortacle: the output refers to vertices not mentioned in the input (No-New); the output duplicates vertices (Uniqueness), and the output's length is not equal to the number of vertices used in the input (Same-Num-Vertices). Given the egregiously incorrect output, we were curious how the predicate managed to slip past all the other test cases.

Examination revealed that, while the student's predicate had been divided into roughly the subproperties we expected, their duplicate-checking code had been turned off. Instead, the code for checking the Sortedness subproperty was written to also implicitly check for duplication in the output, but in a subtly incomplete way. For every vertex in the output, the code was counting the number of incoming input-edges and comparing that number with the number of incoming "edges" represented by the previous output vertices.

After a brief attempt to produce more examples demonstrating this same error, we turned to the strategy in section 3: adding an error property for the underlying issue.

Another 15 minutes of human effort produced error subproperties for both Hypothesis and Alloy representing the sub-check the student was *correctly* performing.

Adding this subproperty, which we called IncomingEdgeCount, enriched the set of buckets for Toposortacle. Most usefully, these included a bucket where the new subproperty was true, but Uniqueness was not. With this new bucket, the student predicate was reliably labeled incorrect by both the Solver and Hypothesis tools.

While this specific error was not reproduced by any other submissions, the exercise gave us confidence that our process for creating and using error properties is effective, even in the presence of quite subtle mistakes. It also illustrates the value of generating multiple tests within a bucket, and leaving some un-shrunk.

## 9 Quantitative Evaluation and Comparison to Prior Results

Before we begin discussing comparisons, we note that there are two distinct prior artifacts to compare against. The first, naturally, is the (mostly) human-generated suites for the same problems from pj21. This is a comparison of the subproperty approach. The other is arguably more intriguing. The problems we study have been used in classes for several years, during which time several generations of course staff curated test suites for the predicates students write. We can ask how well this paper's approaches compare against those manual suites at testing student predicates.

There are, in addition, a few other natural questions: about how the two automated approaches compare against each other, and how well the generation process works. Taken together, we arrive at the following research questions:

- rq1: How do our automatic-generation approaches compare to each other in terms of error-catching power?
- rq2: How do our automatically generated test suites compare to those from pj21?
- rq3: How do our automatically generated test suites compare to multi-year curated human-generated suites for the same problems?
- rq4: For these problems, how much of the power set is concretizable, and how much of it must be concretized to find all errors?
- rq5: How expensive is the automatic process of producing test suites?

We address each of these in turn. Note that we have split questions of performance in two: rq5 focuses on the performance of generating individual suites, while rq4 confronts the scaling of the overall bucketing process. (Recall that we discussed developer effort qualitatively in sections 5 and 6.)

### 9.1 Comparison of Automated Methods and Prior Work (rq1, rq2)

First we address the question of how each of the automated methods did, compared against each other and against the suites from pj21. Our subproperty decompositions are *not* a perfect match to pj21, due to the addition of error properties and the decomposition of properties like Same-Elements into the smaller No-New and None-Dropped. Yet, by following the discipline sketched in section 4, we can identify sets of buckets that roughly correspond to the independent subproperties in pj21.

Doing this requires some care, however. The non-error subproperties in this work are identical to pj21's, with one exception:

pj21's Same-Elements property checks that the elements of the input and output are the same, which allows for mishandling of duplicates. This underconstraint made it possible for the Same-Size property to be independent, but at a cost; pj21's Sortacle properties do not actually imply correctness.

To account for this, we report two separate analyses: one involving the original pj21 version (Same-Eles-Weak) and another, corrected, version that uses multisets (Same-Eles-Strong). This fact, along with error subproperties and decompositions, means that the total number of properties for each problem is not the same between pj21 and the present work. Thus, in this section we will take care to always note when we are discussing pj21's properties.

For each problem, the pj21 subproperties are as follows. **Sortacle:** Same-Size (the input and output lists must be the same length), Same-Elements (the input and output lists contain the same sets of elements), and Ordered (the output list must be ordered correctly). **Matcher:** Stable (no unmatched pair would prefer each other to their match in the output), Uniqueness (no element of the output is matched to more than one counterpart), and Complete (all elements of the input are matched). **Toposortacle:** Same-Elements (for consistency with table 1, we represent this as the conjunction of No-New and None-Dropped), Ordered (the elements of the output are ordered with respect to the input partial order), Uniqueness (the output contains no duplicate vertices), and Same-Num-Vertices (the output contains the same number of vertices as the input mentions).

Beyond reporting on each of these subproperties, we also measure *new* errors found by buckets that were *not focused* on a specific subproperty—buckets that did not correspond to a subproperty in pj21. We group these under the Unfocused label. For Sortacle, we report Unfocused-Weak and Unfocused-Strong, which respectively use either the incorrect or the correct permutation subproperty.

We set both the solver and Hypothesis to produce 10 tests per bucket to keep down the overhead of running the suites for all buckets across each assignment. Our generated suites thus contained 10 positive tests for each problem, and a number of negative tests depending on which buckets were run (taking implications into account). Each method produced 360 negative tests for Toposortacle, 210 for Sortacle, and 430 for Matcher—section 9.3 explains these numbers. For reference, pj21's handcrafted suites contained 9 positive and 18 negative tests for Toposortacle, 32 positive and 39 negative tests for Sortacle, and 4 positive and 3 negative tests for Matcher.

We will eliminate predicates that could not run or were otherwise so deeply flawed as to render a fine-grained analysis unhelpful. pj21 did so [40, section 4.1] by eliminating predicates that did not pass a positive Functional test suite, which involved inputs for which there was only one correct output, or a negative All suite whose tests failed all properties simultaneously. A second positive suite, Relational, containing multiple valid pairs with the same inputs, was *not* used for filtering, but served as an additional point of evaluation.

**Table 2** Comparing (number of errors found by) the **S**olver-based and **H**ypothesis-based approaches against each other and against suites from PJ21. TOTAL rows aggregate across PJ21 subproperties and unfocused suites.

| Problem | Suite Group | PJ21 | S | $\pm(S;21)$ | H | $\pm(H;21)$ | $\pm(H;S)$ |
|---|---|---|---|---|---|---|---|
| Sort. | SAME-SIZE | 16 | 15 | 6;7 | 20 | 10;6 | 5;0 |
| Sort. | ORDERED | 4 | 6 | 4;2 | 8 | 4;0 | 2;0 |
| Sort. | SAME-ELES-WEAK | 34 | 46 | 12;0 | 46 | 14;2 | 2;2 |
| Sort. | SAME-ELES-STRONG | 34 | 72 | 40;2 | 76 | 42;0 | 4;0 |
| Match. | STABLE | 10 | 28 | 18;0 | 32 | 22;0 | 5;1 |
| Match. | UNIQUENESS | 0 | 8 | 8;0 | 9 | 9;0 | 1;0 |
| Match. | COMPLETE | 2 | 78 | 76;0 | 75 | 73;0 | 1;4 |
| Topo. | ORDERED | 11 | 12 | 1;0 | 12 | 2;1 | 1;1 |
| Topo. | NONE-DROPPED | 22 | 22 | 1;1 | 23 | 1;0 | 1;0 |
| Topo. | NO-NEW | 32 | 32 | 0;0 | 32 | 0;0 | 0;0 |
| Topo. | SAME-NUM-VERTICES | 28 | 31 | 3;0 | 31 | 3;0 | 0;0 |
| Topo. | UNIQUENESS | 52 | 53 | 1;0 | 52 | 1;1 | 0;1 |
| Sort. | POSITIVE | 2 | 4 | 3;1 | 11 | 9;0 | 7;0 |
| Match. | POSITIVE | 9 | 6 | 0;3 | 16 | 9;2 | 12;2 |
| Topo. | POSITIVE | 45 | 43 | 1;3 | 46 | 1;0 | 3;0 |
| Sort. | UNFOCUSED-WEAK | 0 | 27 | 27;0 | 28 | 28;0 | 2;1 |
| Sort. | UNFOCUSED-STRONG | 0 | 2 | 2;0 | 0 | 0;0 | 0;2 |
| Match. | UNFOCUSED | 0 | 3 | 3;0 | 7 | 7;0 | 5;1 |
| Topo. | UNFOCUSED | 0 | 9 | 9;0 | 9 | 9;0 | 0;0 |
| Sort. | TOTAL | 50 | 77 | 30;3 | 83 | 33;0 | 6;0 |
| Match. | TOTAL | 14 | 98 | 84;0 | 102 | 90;2 | 7;3 |
| Topo. | TOTAL | 86 | 96 | 11;1 | 96 | 11;1 | 1;1 |

While our automated approaches can generate positive tests, they cannot at present distinguish between functional and relational problem instances, and so for consistency we use the same FUNCTIONAL positive test-suite as PJ21 did to exclude solutions. However, since our goal is to evaluate the test-generation process itself, rather than students' submissions, we also compare 10 positive tests from Hypothesis and the solver against the FUNCTIONAL and RELATIONAL suites from PJ21; we denote this combination POSITIVE. Although our focus is on negative tests produced for each bucket, we wish to avoid potentially concealing any issues with positive test generation.

Table 2 lists, across each problem, how many erroneous predicates were found by each method. Since the errors found by different suites need not be related, the "±" columns report the size of the pairwise set difference: e.g., $\pm(S;H)$ is $|S \setminus H|$ and $|H \setminus S|$ for the Solver-based and Hypothesis-based methods respectively.

**Interpretation: Comparing Automated Methods** The last column compares the SAT-solver and Hypothesis suites. Each does better on different subproperty groups, but with sizable set differences: in some rows, Hypothesis found 6, 7, or 12 errors that SAT did not; in others, SAT found 2 or 4 novel errors. The Hypothesis approach is very effective, but at significant runtime cost (section 9.4), so a two-pronged approach has value. It is also worth noting that the SAT-solver's worst performance seems to be on positive tests (the 12 mentioned was on positive test buckets). While the exact cause is uncertain, this only underscores the value of multiple approaches.

**Interpretation: Comparing Against PJ21** On Sortacle, the solver and Hypothesis made a strong showing, often doubling (or more) the number of errors found. This is especially, albeit unsurprisingly, true on the SAME-ELES-STRONG property variant, as this variant uses the correct SAME-ELEMENTS property that PJ21 underspecified. On Matcher, PJ21 reported *no* failures of UNIQUENESS and only two for COMPLETE, yet automation found 9 and 76 failures.

For UNIQUENESS, the subproperty corresponded to exactly one bucket, since there were no implications between UNIQUENESS and other subproperties. Thus, there was no advantage gained from a finer-grained decomposition. Sadly, however, PJ21's suite had only one test for this subproperty. For COMPLETE, we note that PJ21 likewise only had one test. Furthermore, it did not distinguish between the two distinct input lists: predicates that only checked completeness for *candidates* but not *companies* would be missed by PJ21. These issues underscore again the need to generate multiple tests per bucket, and the value of carefully-crafted automation (i.e., elimination of duplicates) to that end.

On Toposortacle alone are our results somewhat similar to PJ21. The TOTAL row for Toposortacle in table 2 reports an 11;1 set difference for both automated methods; further examination reveals that the solver and Hypothesis both found 1 error the other missed, again highlighting the value of an ensemble approach.

In fact, *the union of the Hypothesis and Solver suites caught all bad predicates that PJ21's suite did*. The results for SAME-SIZE in Sortacle are an artifact of splitting the weak vs. strong SAME-ELEMENTS properties; errors PJ21 caught for SAME-SIZE were caught by automated tests in the SAME-ELES-STRONG bucket instead.

Finally, we observe that in all 3 problems, several new errors were found via tests with multiple subproperty failures (UNFOCUSED). We take this as strong evidence that there is value to a complete power-set enumeration.

### 9.2 Comparison vs. Course-Staff Curated Suites (RQ3)

Table 3 summarizes how our generated suites compare against the staff-curated test suites. Those suites evaluate the whole predicate, not subproperties, so we compare them against all buckets combined. We use "H" for Hypothesis and "S" for SAT.

Here we include *all* predicates, even those that failed the PJ21 FUNCTIONAL suite, to make this a direct comparison between our suites and the curated suites. (When excluding predicates that failed FUNCTIONAL, the proportional results are similar.) Because table 2 already provides the difference in yield between our two approaches, we omit repeating that set difference here. For reference, the staff suites contained 6 positive and 10 negative tests for Toposortacle, 12 positive and 19 negative tests for Sortacle, and 12 positive and 9 negative tests for Matcher.

**Table 3** Comparing against curated assignment suites (by number of errors found).

| Problem | TA Suite | S | $\pm(S;TA)$ | H | $\pm(H;TA)$ |
|---|---|---|---|---|---|
| Sortacle | 82 | 79 | 3;6 | 86 | 5;1 |
| Matcher | 101 | 121 | 22;2 | 125 | 26;2 |
| Toposort | 135 | 159 | 27;3 | 159 | 27;3 |

**Interpretation** Our new method beats the staff suites on all problems, but often not by much, and the manual suites still find some errors that automation misses. However, it is worth noting that the manual suites have the advantage of being revisited several times by different people with the benefit of experience and hindsight, and with the immediate stakes of giving in-flow student feedback, rather than being a freshly created, first-cut evaluation of a new technique.

For Matcher, our improvements were somewhat scattershot across all buckets, which would seem to indicate that the TA suite needs more diverse test cases across the board. There was exactly 1 error in both Sortacle and Matcher, and 2 errors in Toposortacle, that the TA suite found but neither the solver nor Hypothesis did. For Toposortacle, closer investigation revealed a difference in input type assumptions: the TA suite used arbitrary strings as node names in Toposortacle, whereas our generators used only integers. For the others, one was from a *positive* TA test (which was not the focus of this work; we only generated 10 positive tests per method as a sanity check), and the only other came from a TA test with an intricate weaving of input list elements; this will become a new error subproperty in the future.

The strong improvement on Toposortacle can be attributed to the fact that the TA suite, for all its careful design, actually neglected to include any tests where elements of the output were not unique, but did not also fail to appear in the input. That is, the TA suite had no discriminatory power over a predicate that checked No-New but not Uniqueness (or vice versa). This error was therefore found purely because of the subproperty decomposition approach, confirming the technique's power to reveal subtle oversights that even experts can fall prey to.

### 9.3 Scaling the Power Set (RQ4)

Once all error subproperties are included, the present work had 6 subproperties per problem. A decomposition of 6 subproperties theoretically induces a power set of $2^6 = 64$ buckets. However, when subproperties are no longer independent, a number of the buckets are logically contradictory due to implications between subproperties. Of the 64 total potential buckets for each problem, Sortacle had only 22 that were concretizable, Matcher had 44, and Toposortacle had 37. As new properties are added, satisfying the new property will only be possible in an increasingly smaller subset of the currently existing buckets. In practice, this mitigates the scaling impact as the number of subproperties increases.

**Table 4** Statistics on runtime (in seconds) for test generation across all buckets, rounded to the nearest tenth of a second, for the **S**olver and **H**ypothesis approaches.

| Problem | N | S Min | S Avg | S Max | S StD | H Min | H Avg | H Max | H StD |
|---|---|---|---|---|---|---|---|---|---|
| Sortacle | 1 | 0.7 | 0.8 | 1.5 | 0.2 | 0.0 | 1.1 | 5.2 | 1.4 |
| Matcher | 1 | 2.7 | 3.4 | 5.6 | 0.7 | 0.1 | 9.5 | 111.4 | 20.3 |
| Toposrt. | 1 | 1.4 | 1.7 | 2.4 | 0.3 | 0.0 | 1.3 | 9.5 | 1.7 |
| Sortacle | 10 | 0.7 | 0.9 | 1.5 | 0.2 | 0.4 | 88.4 | 804.6 | 182.0 |
| Matcher | 10 | 3.9 | 5.1 | 7.2 | 0.8 | 2.4 | 191.2 | 3684.2 | 574.0 |
| Toposrt. | 10 | 1.6 | 2.1 | 2.9 | 0.3 | 0.6 | 21.1 | 170.0 | 35.7 |
| Sortacle | x | 0.7 | 1.0 | 3.1 | 0.4 | – | – | – | – |
| Matcher | x | 2.6 | 3.5 | 4.3 | 0.5 | – | – | – | – |
| Toposrt. | x | 1.9 | 17.7 | 139.8 | 32.2 | – | – | – | – |

### 9.4 Test-Generation Runtime (RQ5)

Finally, we examine the different runtime characteristics of each method. Table 4 reports, across all buckets in the power set of each problem's decomposition, the maximum, minimum, mean, and standard deviation of runtime (in seconds) for test generation. We report separate numbers for: generating the first test, generating 10 tests, and inferring that a bucket cannot be concretized (labeled "x")—a task that we left solely to SAT, since it was able to exhaustively explore all pairs (up to the bounds given). Since Hypothesis could not deduce a bucket was contradictory (we terminated a bucket after 20,000 unproductive candidates) we leave those entries as "–".

All measurements were done on a 2017 MacBook Pro (i5 2.3 GHz, 8 GB RAM); only one core was used by test generation. We ran SAT-based and Hypothesis generation via Racket v7.9 and pypy3 (a fast implementation of Python 3) v7.3.4 respectively.

**Interpretation** The maximum times for Hypothesis were higher across all buckets, sometimes by multiple orders of magnitude. Hypothesis also shows universally higher average generation times, as well as wider variance. This is true even for the first test produced, even though we allowed Hypothesis to return a small example (section 5, I-O Pair Generator) first in every bucket, yet the solver will not necessarily do so.

Hypothesis did, however, surpass or achieve parity with the solver in the minimum case; Hypothesis found tests more quickly than the solver on *some* buckets. We believe this is because the solver relies on a boolean SAT engine, and so it must pay an up-front cost (per bucket) to translate its relational formulas into purely boolean constraints. In contrast, Hypothesis can immediately start to generate test candidates.

Having paid this initial cost, the solver is *incremental*: its search state is saved between answers, so that previous work is not repeated. We see this effect strongly: e.g., the minimum cost to create 1 test vs. 10 tests for Sortacle differed by less than rounding error (we round both to $0.7s$). Hypothesis does not exhibit this phenomenon.

It is interesting that Matcher proved most challenging for both approaches on concretizable buckets, yet Toposortacle was instead the most challenging for the solver when it came to proving buckets were contradictory; it spent 139.8 seconds at maximum per contradictory bucket, and 17.7 seconds on average. The standard deviation of 32.2 seconds shows the impact of a few particularly challenging buckets on the overall figures. The other two problems were tame by comparison.

## 10 Other Related Work

While PBT remains uncommon in undergraduate curricula, some (e.g., Fredlund, Herranz and Mariño [9], Scharff and Sok [30], Dadeau and Tissot [5], and Wikström [39]) use PBT in computing education or endeavor to motivate PBT. To our knowledge, our work is unique in its focus on evaluating student-authored PBT predicates.

QuickCheck [3], the original PBT framework, provides a library for random input generation and a testing harness that detects when outputs violate stated correctness properties. The PBT framework we employ, Hypothesis [17], is similar.

Solvers have long been widely used to generate tests. Some, such as Korat [23], TestEra [21], and Whispec [32] even use the Alloy toolchain. Concolic software testing (e.g., CUTE [31]) also rely on solver technology. Jahangirova, Clark, Harman, and Tonella [13, 14] use mutation and test-generation techniques to improve test oracles (largely assert statements) in existing code. Iorek [27] uses SMT solvers to produce test suites with a high *difference metric* between tests. Since the focus of this work is on testing PBT predicates themselves and obtaining semantic insight in the process, all these tools might have made admirable additional paths for test generation, but they are orthogonal and complementary to this paper. Existing coverage and mutation-based [7, 26] techniques, or other types of test amplification [6] likewise offer an orthogonal, but promising, avenue for future work. Work aimed at applying these ideas to Alloy itself, e.g., the AUnit [33, 34] framework, might be especially applicable, since they are designed to work on the specifications themselves.

## 11 Discussion and Conclusion

We believe that this work puts subproperty decomposition on sound, scalable, and useful footing—at least for educational purposes. We hope that, if PJ21 [40] stimulated interest in teaching PBT, this work will provide a solid footing, techniques, and tooling to ease translating that interest into curricular practice. To that end, we have made our scripts and a full Toposortacle example available in an online appendix.[1]

At the very least, subproperty decomposition is clearly a feasible way to augment existing test suites for PBT. We say this with some chagrin; humans (even ones trained in formal methods and PBT) may not consider subtle yet important combinations of subproperties (as in the Toposortacle suite). Even in the absence of automation, the discipline of breaking correctness down and thinking about tests per bucket, rather than per subproperty in isolation, can bear fruit.

[1] http://cs.brown.edu/research/plt/dl/pj2021/

**From Errors to Misconceptions** Since we were working with prior-year data, all this paper's analyses are after-the-fact, rather than integrated into each course. By itself, our approach only finds errors and isolates their semantic nature; it cannot distinguish true misconceptions from bugs. Providing real-time feedback could yield deeper insight. It may not always be reasonable to make judgments even then, but this work provides a place to start. Without undertaking it, we would never have realized how common some errors were, and applying this discipline will likely yield new insights each year.

**Fingerprinting Errors and Error Properties** How semantically different, up to buckets, are students' incorrect predicates? For each predicate, we computed a *fingerprint*: the set of buckets on which the predicate failed. We initially expected only a few fingerprints, perhaps corresponding to specific misconceptions. Instead, we found more fingerprints than we can detail here: 38 for Sortacle, 102 for Matcher and 75 for Toposortacle. In hindsight, this is reasonable: even if two students possess the same high-level misunderstanding, they may not reflect it identically in their code. We stress that subproperties, especially error subproperties, still identify meaningful patterns within the fingerprints; our approach still provides useful insight. Nevertheless, it seems that (to misquote Tolstoy [35]) while correct predicates are semantically identical, incorrect predicates are often wrong in their own unique way. This suggests an area for future work: which differences arise from gaps in understanding (and thus yield new error subproperties), rather than variation in code structure?

**Exception Handling** When analyzing our data, we considered a predicate throwing an exception to be equivalent to a wrong answer. But this may hide a deeper curricular issue. Should PBT assignment specifications say not only when predicates should return true, but also to throw exceptions only when input preconditions are violated? This could lead to an exciting investigation of students' assumptions about preconditions.

**Student Performance** Although our main goal was to put PJ21 on better technical footing, our analysis discovered places, like the COMPLETE subproperty for Matcher, where students had more trouble than PJ21 detected. As a result, student performance is a more mixed story: they still did well overall, but there were errors not previously caught. This does not alter our perception that PBT is accessible for undergraduates.

**Acknowledgements** We thank the anonymous reviewers, especially Reviewer C, for their careful reading and detailed comments. This work was partly supported by the US National Science Foundation. This research was also developed with funding from the Defense Advanced Research Projects Agency (DARPA) and the Air Force Research Laboratory (AFRL). The views, opinions and/or findings expressed are those of the authors and should not be interpreted as representing the official views or policies of the Department of Defense or the U.S. Government.

## About the authors

**Tim Nelson** (tbn@brown.edu) preaches the good news of logic and computing at Brown University.

**Elijah Rivera** Elijah Rivera (elijah_rivera@brown.edu) is a PhD student focused on leading students to water *and* making them drink at Brown University.

**Sam Soucie** Sam Soucie (samdavidsoucie@gmail.com) is an Indiana University alumnus, former research assistant at Brown University, and currently works in the financial technology industry in Chicago, Illinois.

**Thomas Del Vecchio** Thomas Del Vecchio (thomas_del_vecchio@brown.edu) is a Lambda-calc lovin' Pyret studying CS as an undergrad at Brown U.

**John Wrenn** (jswrenn@cs.brown.edu) is a guerilla archivist, tandem bicycle evangelist, and PhD student at Brown University.

**Shriram Krishnamurthi** (shriram@brown.edu) is the Vice President of Programming Languages (no, not really) at Brown University.